\documentstyle[prl,aps,multicol,epsfig,epsf]{revtex}

\large 
\begin{document}
\draft 
\title{Andreev reflection off a fluctuating
superconductor in the absence of equilibrium} 
\author{P. Devillard$^{1,2}$, R. Guyon$^{1,3}$, T.
Martin$^{1,3}$,  I. Safi$^{1,4}$ and B. K. Chakraverty$^{5}$} 
  
\address{ $^1$ Centre de Physique 
Th\'eorique,  
 Case 907 Luminy, 13288 Marseille Cedex 9, France} 
\address{$^2$ Universit\'e de Provence, 13331 Marseille Cedex 
03, France}  
\address{$^3$ Universit\'e de la M\'editerran\'ee,  
13288 Marseille Cedex 9, France}
\address{$^4$ Laboratoire de Physique des Solides,
Universit\'e de Paris-Sud, 91405 Orsay}
\address{$^{5}$ Laboratoire d'Etudes des Propri\'et\'es
Electroniques des Solides, CNRS, BP 166, 38042 Grenoble}

\maketitle 
\begin{abstract} 
Andreev reflection between a normal metal and a
superconductor  whose order parameter exhibits quantum phase
fluctuations is examined. The approach chosen is non
perturbative in the tunneling Hamiltonian, and enables 
to probe the whole range  of voltage
biases up to the gap amplitude. 
Results are illustrated using the one--dimensional Josephson
junction array model previously introduced in the linear
response regime. Phase fluctuations are shown to affect the
differential conductance and  are compared to the result of
Blonder, Tinkham and Klapwijk for a rigid BCS superconductor. 
The noise spectrum of the Andreev current is
also obtained and its second derivative with respect to frequency
is proposed as a direct tool to analyze the phase fluctuations.
\end{abstract}

\section{Introduction}

In the last decades, a considerable effort has been devoted towards the 
study of the transport properties of  
normal metal-superconductor (NS) junctions.
The situation for the Andreev current and finite frequency noise is well 
understood when the superconductor
is of the BCS type. 
More recent works have dealt with superconductors whose order 
parameter has a d-wave symmetry \cite{bruder,tanaka}. 
The role of collective
modes arising from the phase of the fluctuations of the superconductor 
has also been addressed in the framework of linear response theory in 
high-$T_c$ materials \cite{kim}, as well as in
one-dimensional array of Josephson junctions  
\cite{falci}.
Some recent attempts have also included the effect of classical
phase fluctuations of the order parameter on Andreev transport,
either in the tunneling regime \cite{sheehy} or using Bogolubov-de
Gennes equations \cite{choi}. 

\begin{figure}  
\epsfxsize 8 cm  
\centerline{\epsffile{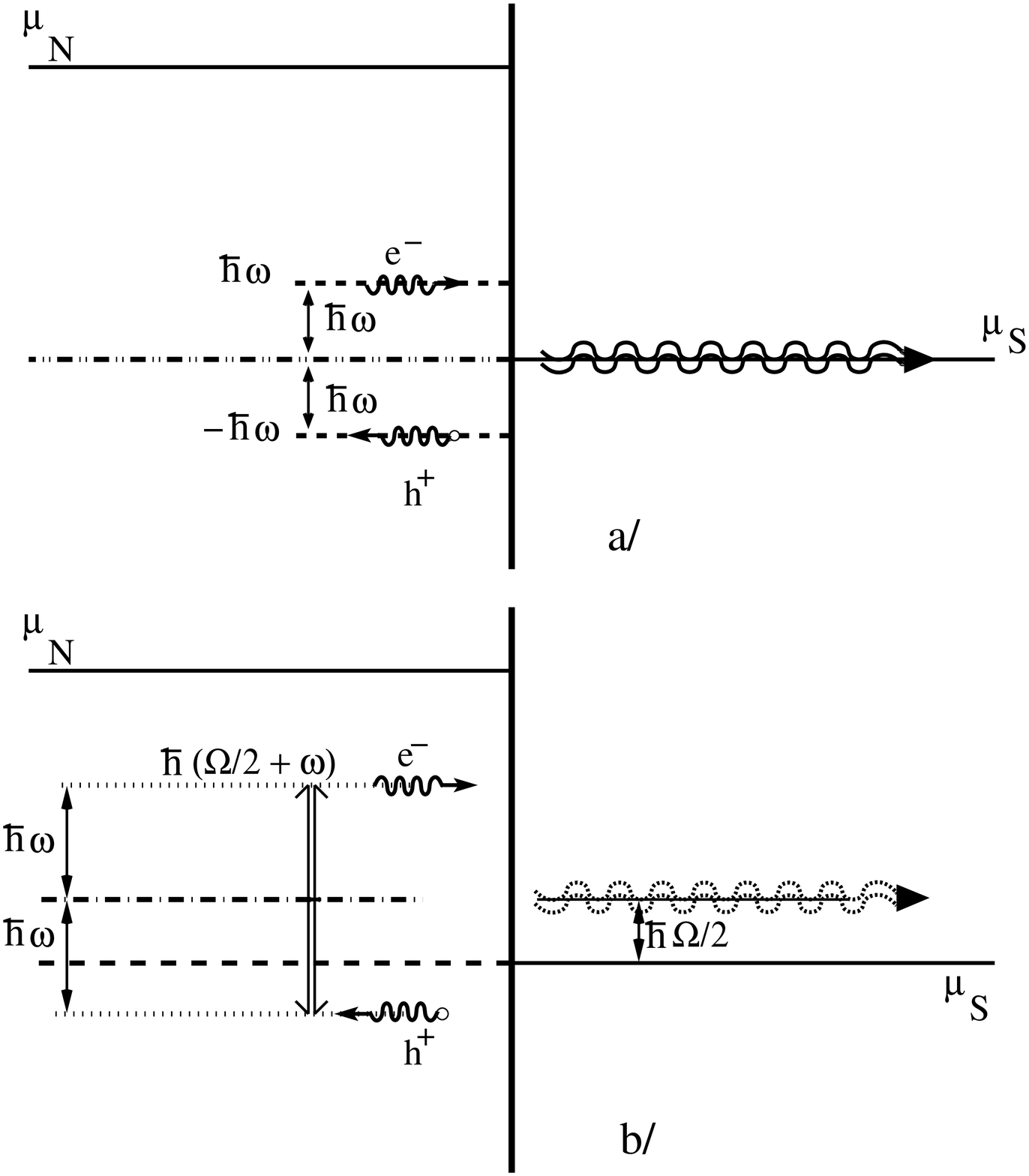}}  
\medskip  
\caption{\label{rodolphe1}
Andreev reflection (bias voltage $eV=\mu_N-\mu_S$): 
a) In a rigid superconductor, the two
electrons 
 (one electron $e^-$ and a time reversed hole $h^+$)
have opposite energies $\pm\hbar\omega$ 
 with respect to the
superconductor chemical potential $\mu_S$,  and a Cooper pair is
created on the right hand side. 
 b) In the presence of 
phase fluctuations, the superconductor absorbs a collective
mode with energy  $\hbar\Omega$, so that the energies of the
two incoming electrons 
 are no more symmetrical with respect to
$\mu_S$.}   
\end{figure} 

A physical picture of the Andreev process is drawn in Fig. \ref{rodolphe1}. 
A voltage bias is applied, such that the chemical potential of 
the metal $\mu_N$ lies $eV$ above that of the superconductor
$\mu_S$. The origin of the energies is taken at $\mu_S$. 
In Andreev reflection with conventional, rigid
superconductors, Fig. \ref{rodolphe1}a, 
an electron below $\mu_N$, with an energy $ \hbar \omega$
is reflected as a hole and a Cooper pair propagates in the superconductor.
This pair has zero energy. The reflected hole can be interpreted as
an electron below $\mu_S$, with energy $ -\hbar \omega$, which enters
the superconductor.
At zero temperature, the only possible values of $\hbar\omega$ are
between $- e V$ and $eV$, see Fig. \ref{rodolphe1}a , because both electrons states must
have an energy below $\mu_N$. 

Turning to the case of a superconductor with
 phase fluctuations, Fig. \ref{rodolphe1}b:
exciting a collective mode in 
the superconductor costs an energy $\hbar \Omega$. 
One can therefore take an electron of energy $\hbar(\Omega/2 + \omega)$
and reflect a hole, which is equivalent to a missing electron 
of energy$ - \hbar (\omega - \Omega/2)$.
The allowed values of $\hbar \omega$ must be 
smaller than $eV - \hbar \Omega/2$ 
(no electrons present above $\mu_N$). Note that here, one has to 
sum over all possible values of $\hbar \Omega\in[0;2eV]$, taking care of the fact  that this is a weighted sum, with a probability distribution
$P(\Omega)$ which is determined by the Hamiltonian describing 
the phase.

In this paper, the Andreev current is calculated non-perturbatively 
in the tunneling Hamiltonian for N-S junctions whose
Hamiltonian includes quantum phase fluctuations. 
The approach is inspired from the Keldysh technique of 
references \cite{caroli,martin-rodero}, and enables
the calculation of the current-voltage characteristics 
of the Andreev current when 
the bias voltage $eV$ is close to (but smaller than) 
the gap amplitude $\mid\!\Delta\!\mid$. 
In particular, 
this corresponds to the experimental situation encountered in
scanning tunneling microscope (STM)  experiments performed on
high-$T_c$ materials \cite{deutscher,gonnelli,mourachkine}. 
Although the calculation is restricted here to an s-wave order parameter,
qualitative features can be derived for transport in N-S
junctions for arbitrary superconductors, provided that there is a gap.
Here, a phase-only effective Hamiltonian will be used, which was 
previously derived in the
literature \cite{De Palo}. In one dimension, this model is isomorphic to an 
array of Josephson junctions, with Josephson coupling 
energy $E_J$ and charging energy $E_0$.  
The current and the finite frequency noise spectrum 
$S(\omega)$ are both derived, which requires   to go beyond linear
response theory. In particular, it will be shown that the 
second derivative of the noise with respect
to the frequency $d^2 S / d \omega^2 $ gives a direct access
to information about phase fluctuations, independently of the model
chosen to describe the latter.  
 
There are other situations where Andreev scattering is mediated 
by excitations at the normal supraconductor boundary. 
If the normal side is replaced by a ferromagnet \cite{falko},
the spin of electrons pairs emerging from/entering the 
superconductor can suffer a spin flip accompanied by the
emission/destruction of a magnon. This additional scattering
channel effectively may enhance the Andreev current.
Except for the case of a half-metal, electrons with opposite
spins as well as electrons with the same spin then
contribute to the Andreev current.  The  differences with our
phase fluctuations calculation will addressed below.

The paper is organized as follows: in section \ref{section2} , the model
Hamiltonian is discussed. In section \ref{section3},
 the perturbative scheme which 
allows to derive the relevant Green's functions for the 
transport properties is described. In section \ref{section4}, the phase 
fluctuations are introduced in this framework. The 
current-voltage characteristics associated with the linear
Josephson junction array model of Ref. \cite{falci}
for the phase appears in section \ref{section5}, as a function of 
the physical parameters $E_J$, $E_0$, as well as the 
barrier transparency.

\section{Model Hamiltonian}

\label{section2}

The Hamiltonian describing the NS junction is composed
of three terms: $
H = H_N + H_t + H_{JJA}   
$.
The first term is simply the Hamiltonian for the metal
which is specified to a one band model
$
H_N  =  \sum_{k , \sigma} \epsilon_{k , \sigma} 
c^{\dagger}_{k , \sigma} c_{k , \sigma}
$
with width $W$. 
The second term $H_t$ is the tunneling term,
\begin{equation}
H_t  = 
 \sum_{N,S,\sigma} \Gamma  c_{N,\sigma} c_{S , \sigma}^{\dagger} + h.c.,
\label{5}
\end{equation}
where $\Gamma$ is a hopping amplitude which transfers electrons
to/from the superconductor.

In the superconductor, it is assumed that the low
energy lying excitations are given  by fluctuations of the phase
of the order parameter. 
If we forget the poorly screened long-range part of the Coulomb interaction, 
the Hamiltonian is the same
as the one describing an array of Josephson junctions \cite{falci,glazman}.
\begin{equation}
H_{JJA}  =  \sum_{i,j} E_J  \cos(\theta_i - \theta_j) 
 +    \sum_i {q_i^2 \over 2C}~,  \label{4}
\end{equation}
where
$E_J$ is the Josephson coupling (between islands $i$ and $j$), 
 the operator
$\theta_i$  is the phase of the order parameter at ``site'' $i$,
$q_i$ is the charge on island $i$ and
${q_i^2 \over 2 C}$ is the charging energy of island $i$.
Eq. (\ref{4}) can be put on a firmer basis, starting from
microscopic models. In particular, a phase-only Hamiltonian
$H_S$ has been derived from  various microscopic models such as
the attractive Hubbard model  \cite{De Palo}:
\begin{equation}
H_S  =   \sum_{i,j} {\hbar^2 N_S^0 \over 2 m^* } 
 \cos(\theta_i-\theta_j)   +  \sum_i {1 \over 2 } 
{m^* v^2 \over N_S^0} \rho_i^2
  + \sum_{i,j} {\pi e^2 \over \varepsilon}  
{\rho_i \rho_j \over \mid\!r_i-r_j\!\mid}~,
\label{3}
\end{equation}
where $N_S^0$ is the bare superfluid density,  $\varepsilon$ is the
dielectric constant, $\mid\! r_i - r_j \!\mid$ is the  distance
between site $i$ and $j$. The first term is the Josephson
coupling between  location $i$ and location $j$. 
The second term is the local density fluctuation energy 
 $(\rho - \langle\rho\rangle)^2/2\chi_0$, 
 where $\langle \rho\rangle$ is the average density 
and $\chi_0$ a density susceptibility. 
The last term is due to poorly screened long range Coulomb interactions. 
Since electron and hole-quasiparticles are not quasiparticles 
 for the superconductor, the injected electrons and holes will 
 decay and form pairs that will excite the eigenmodes of 
$H_S$.

\section{Non--equilibrium Green's functions and transport}

\label{section3}

A voltage bias is introduced between the normal metal and the superconductor
and the average current  reads
\begin{equation}
I(t) = {i e \over \hbar}  \sum_{\sigma}
 \Gamma  \langle c_{N,\sigma}^{\dagger}(t) 
 c_{S,\sigma}(t)  \rangle -  h.c. .
  \label{6}
\end{equation}
Tunneling occurs from the bulk
of the normal metal to the bulk of the superconductor through a narrow
constriction, the tip of a STM for example.

The Keldysh formalism is used to express the current in terms of
Green's functions. The definitions and properties of this system
have been studied in detail in Ref. \cite{martin-rodero}. Here
we adopt the same notations and start from their expression:
\begin{equation} 
\langle I(t) \rangle  =  {2 e \over \hbar^2} 
\mid\!\Gamma\!\mid^2   \int_{-\infty}^{\infty}  \bigl\lbrack 
g_{NN11}^{+-}(t,t_1)G_{SS11}^{-+}(t_1,t) - g_{NN11}^{-+}(t,t_1)
G_{SS11}^{+-}(t_1,t)  \bigr\rbrack 
dt_1. \label{8} 
\end{equation}
The dressed  Keldysh Green matrices on the superconducting side
are expanded using the Dyson expansion  
\begin{equation}
\hat{G}_{SS}^{\pm\mp} (t,t')= \left[
\hat{G}_{SS}^{r}\hat{{\Sigma}}^{\dagger} \hat{g}_{NN}^{\pm\mp}
\hat{{\Sigma}} \hat{G}_{SS}^{a}\right](t,t'),  
\label{9}
\end{equation}
where $\hat{\Sigma}$ is the self energy matrix describing the
hopping from the normal side to the superconductor, and 
$\hat{\Sigma}^\dagger$ describes the opposite process. For
simplicity, integrals over time have been omitted.
Note that only subgap tunneling is considered here, allowing 
to set $\hat{g}_{SS}^{\pm\mp}=0$. 
Inserting the Dyson expansion, the current reads:
\begin{equation}
\begin{array}{l}
{\displaystyle \langle I(t) \rangle = {2 e \over \hbar^4} |\Gamma|^4 
  \int_{\!-\infty}^{\infty}\!\int_{\!-\infty}^{\infty}
\!\int_{\!-\infty}^{\infty}
G_{SS12}^{r}(t_1,t_2)   G_{SS21}^{a}(t_3,t) 
   \times} \\ 
{\displaystyle  \times 
\Bigl( g_{NN11}^{+-}(t,t_1)   
g_{NN22}^{-+}(t_2,t_3)   -  
 g_{NN11}^{-+}(t,t_1)   
g_{NN22}^{+-}(t_2,t_3) \Bigr) 
dt_1 dt_2 dt_3~.} 
\end{array}  \label{current_K}
\end{equation}
The dressed, retarded Green's function on the superconducting
side is given by:  
\begin{equation}
G_{SS12}^r(t^{\prime},t) =
 g_{SS12}^r(t^{\prime},t)   +  
\int_{\!-\infty}^{\infty}\!\int_{\!-\infty}^{\infty} 
\left( \hat{g}_{SS}^r(t^{\prime},t_1)    \hat{\Sigma}_S^r(t_1,t_2)  
 \hat{G}_{SS}^r(t_2,t) \right)_{12} dt_1 dt_2 +
\ldots, \label{13} \end{equation}
with the dressed self energy $\hat{\Sigma}_S^r$ defined as:
\begin{equation}
\hat{\Sigma}_S^r (t_1,t_2)=  \int_{\!-\infty}^{\infty}\!
\int_{\!-\infty}^{\infty} 
\hat{\Sigma}^*(t_1,t^{\prime}_1)  
\hat{g}_{NN}^r(t^{\prime}_1,t^{\prime}_2)
\hat{\Sigma}(t^{\prime}_2,t_2) dt^{\prime}_1 
dt^{\prime}_2~, \label{14}
\end{equation}
and the unperturbed momentum averaged Green's function
in the superconductor reads : 
\begin{equation}
g_{SS12}^r(\omega)   = {\hbar \over W}   
{\mid\!\Delta\!\mid \over \sqrt{\mid\!\Delta\!\mid^2 - 
(\hbar \omega+i \alpha)^2}} 
e^{i \theta_0}  \equiv  \tilde{g}_{SS12}^r(\omega)  
e^{i \theta_0}, \label{7}
\end{equation}
where $\alpha$ is an infinitesimally small positive quantity and 
$\theta_0$ the phase of the superconductor, which appears also in
the gap parameter: $
\Delta  =  \mid\!\Delta\!\mid e^{i \theta_0}
$.

\section{Transport in the presence of phase fluctuations}
\label{section4}

In the presence of phase fluctuations, the Green's function on the
superconducting side contains two time scales \cite{ketterson}. First, 
fast time scales (of the order $\hbar/\Delta$) which were  
described in the previous section in Eq. (\ref{7}). Secondly,
``slow'' time scales associated with the fluctuating phase
$\theta(x,t)$:     
\begin{equation}
\Delta  =  \mid\!\Delta\!\mid  e^{i \theta(x,t)}. \label{16}
\end{equation}
The phase of the order parameter has both spatial 
and temporal variations and is no longer equal
 to $\theta_0  =  \theta(0,0)$. Tunneling is supposed to
 occur close to the interface. The typical 
dimension parallel to the interface
 will be small with respect to both the classical  
 correlation length $\xi^c_{\phi}$ \cite{bkt} and to the
amplitude correlation  length $\xi_a$. In high-$T_c$
superconductors,   $\xi^c_{\phi}$ and $\xi_a$ are of the order
of a few nanometers.  $\xi^c_{\phi}$ is believed to be
generally not much larger than $\xi_a$,   and $\xi_a$ is much
smaller than in BCS superconductors.  We shall thus replace $x$
by zero.  The temporal variations of $\theta$ are due to
quantum  fluctuations; the phase of the order parameter and the
number  of pairs are conjugated variables. These variations
occur on  time scales of the order at least of the plasma
frequency  $\omega_p\equiv\sqrt{8 E_0 E_J}/\hbar$ with
$
E_0=q_0^2 /2 C
$
and $E_J$ appear in the definition of the Hamiltonian of Eq. (\ref{4}).
Equivalently, $\hbar \omega_p < \mid\!\Delta\!\mid$, which is the contrary 
to what usually
occurs in BCS superconductors. The dynamic nature of the phase
is reflected in the anomalous Green's functions \cite{sheehy}: 
\begin{equation} 
g_{SS12}^{r,a}(t^{\prime},t)  =   
\tilde{g}_{SS12}^{r,a}(t^{\prime},t)   e^{i {1 \over 2}  \bigl(
\theta(0,t) + \theta(0,t^{\prime}) \bigr)}  
\label{include_phase} \end{equation}
where $\tilde{g}_{SS12}^{r,a}(t^{\prime},t)$ is the
inverse Fourier transform of $ \tilde{g}_{SS12}^{r,a}
(\omega)$, and $g_{SS21}^{r,a}(t^{\prime},t)$ has the opposite phase. 
Because of the slow temporal variation of the phase,
the exponential in Eq. (\ref{include_phase}): 
$ \exp[ i(\theta(0,t) + \theta(0,t^{\prime}))/2]\approx 
\exp[i\theta(0 ,t^{\prime})]$ without affecting the results in
a significant manner. 

\subsection{Current and differential conductance}

Substituting Eq. (\ref{include_phase}) in the
current, the fourth (lowest non vanishing) order term in $\Gamma$
which  contributes to the current  is: 
\begin{equation}
\begin{array}{l}
{\displaystyle \langle I(t) \rangle
 =   {2 e \over \hbar^4}  \mid\!\Gamma\!\mid^4  
\int_{\!-\infty}^{\infty}\!
\int_{\!-\infty}^{\infty}
\!\int_{\!-\infty}^{\infty}
\tilde{g}_{SS21}^a(t_3,t)    \tilde{g}_{SS12}^r(t_1,t_2)    
\langle e^{i \theta(0,t_1)}   e^{-i
\theta(0,t_3)}  \rangle  } \\ {\displaystyle  \times  
\Bigl( g_{NN11}^{+-}(t,t_1)     g_{NN22}^{-+}(t_2,t_3)
 -    g_{NN11}^{-+}(t,t_1)  
g_{NN22}^{+-}(t_2,t_3) \Bigr)    dt_1  dt_2  dt_3.} 
\end{array} \label{20}
\end{equation}
The bracket $\langle ...\rangle$ signifies that an average has
been performed over the dynamical degrees of freedom of the
phase. Depending on the nature of the phase, this
will correspond to a Gaussian model -- the so called ordered
phase -- or to a nonlinear model \cite{heinekamp} (see below).  
Note that this expression is analogous to the one derived 
several decades ago by Combescot {\it et al.} for inelastic tunneling
\cite{combescot}:
there the working assumption was that the inelastic coupling did not 
couple directly the two electrodes. Here, this same assumption 
is natural because the phase fluctuations happen only on the 
superconducting side. 
The correlator of the phase and its Fourier transform
are denoted:
\begin{mathletters}
\begin{eqnarray}
p(t)  &=&  \langle e^{i \theta(0 ,  t)}
e^{-i \theta(0 ,  0)} \rangle~, \label{phase_correlator_a}
\\
P(\omega)  &=& {1 \over 2 \pi}  \int_{- \infty}^{\infty}
e^{i \omega t}  p(t)  dt~. \label{phase_correlator}
\end{eqnarray}
\end{mathletters}
$P(\omega)$ can also be viewed as the occupation probability 
of the collective phase modes with energy $\hbar \omega$.
Using the Fourier representation, the current becomes:
\begin{equation}
\begin{array}{l}
{\displaystyle 
\langle I(t) \rangle =   {2e \over h \hbar^3} 
\mid\!\Gamma\!\mid^4 \int_{-\infty}^{\infty} \!
\int_{-\infty}^{\infty} \tilde{g}_{SS21}^{a}(\omega+\Omega/2)   
\tilde{g}_{SS12}^{r}(\omega-
\Omega/2) } \\ {\displaystyle  \times  \Bigl\lbrack 
g_{NN11}^{+-}(\omega + \Omega/2) 
g_{NN22}^{-+}(\omega -\Omega/2)   -
g_{NN11}^{-+}(\omega + \Omega/2) 
g_{NN22}^{+-}(\omega -\Omega/2 )  
\Bigr\rbrack  
P(\Omega) 
d \Omega  d \omega.}
\end{array} \label{23}
\end{equation} 
Next, for frequencies below the gap, one can approximate
$\tilde{g}_{SSij}^{a,r}(\omega\pm \Omega/2)
\simeq \tilde{g}_{SSij}^{a,r}(\omega) \simeq    
\hbar /\left( W  \sqrt{1-\hbar^2 \omega^2/|\Delta|^2}\right)$
($W$ is the bandwidth),   and the current becomes: 
\begin{equation}
\begin{array}{l}
{\displaystyle \langle I(t) \rangle =   
{2e \over h \hbar^3}  \mid\!\Gamma\!\mid^4       
\int_{-\infty}^{\infty} \! \int_{-\infty}^{\infty} 
\tilde{g}_{SS21}^{a}(\omega)  
\tilde{g}_{SS12}^{r}(\omega)   } \\
{\displaystyle  \times 
\Bigl\lbrack 
g_{NN11}^{+-}(\omega + \Omega/2) 
g_{NN22}^{-+}(\omega - \Omega/2)   -
g_{NN11}^{-+}(\omega + \Omega/2) 
g_{NN22}^{+-}(\omega - \Omega/2)  
\Bigr\rbrack  
P(\Omega) 
d \Omega  d \omega,}
\end{array} \label{24}
\end{equation} 
where the two causal Green function elements in Eq. (\ref{24}) 
are related to the electron density of states $\rho(\hbar \omega)$:
\begin{mathletters}
\begin{eqnarray}
g_{NN11}^{+-}(\omega)  &=& 
 2 \pi i  \hat{\rho}_{11}(\hbar \omega-eV)  n_F(\hbar
\omega-eV)~, \label{26} \\
g_{NN22}^{-+}(\omega)  &=& 
 2 \pi  i  \hat{\rho}_{22}(\hbar \omega+eV) 
 \bigl\lbrack n_F(\hbar \omega+eV) -1 \bigr\rbrack ~, \label{27}
\end{eqnarray}
\end{mathletters}
with
$
\hat{\rho}_{11}(\hbar \omega-eV)   \equiv 
\hat{\rho}_{22}(-\hbar \omega+eV) \equiv  
\rho(\hbar \omega)  \simeq \hbar /\pi W $.  

Using the property 
$
\tilde{g}_{SS21}^{a}(\omega)   =  
[ \tilde{g}_{SS12}^{r}(\omega)]^{*} 
$,
the current in the absence of phase fluctuation is recovered by setting 
$P(\Omega)$ to be a delta function:
\begin{equation}
\langle I \rangle_0  = 
{8  e \over h \hbar}  {\mid\!\Gamma\!\mid^4\over  W^2} \int_{-
\infty}^{\infty} \mid \tilde{g}_{SS21}^{a}(\omega)
\mid^{2}  \Bigl\lbrack n_F(\hbar \omega - e V) -
n_F (\hbar \omega+eV) \Bigr\rbrack d \omega, \label{29}
\end{equation}
where subscript $0$ means that in this limit $E_0/E_J  = 0$.
The known result for the Andreev current is recovered 
\cite{martin-rodero}.
It is well known however that even in the absence of phase fluctuations, 
a perturbative calculation to fourth order in $\Gamma$ is not sufficient
to give a satisfactory answer for the whole range of voltage biases between
$0$ and $\Delta$. Indeed, a resummation procedure, explicited by 
Eq. (\ref{13}),  
has to be carried out.
In the absence of phase fluctuations, 
it is thus sufficient to replace $\tilde{g}_{SS21}^{r}(\omega)$ by 
$G_{SS21}^{r}(\omega)$: this allows to recover
the scattering theory results of BTK \cite{BTK}.
 
Turning back to the lowest order contribution of the current
in the presence of phase fluctuations, one obtains:
\begin{equation}
\begin{array}{l}
{\displaystyle \langle I \rangle  =  {8 e \over h\hbar}  
{\mid\!\Gamma\!\mid^4\over W^2}
\int_{-\infty}^{\infty}\!\!\int_{- \infty}^{\infty} 
G_{SS21}^{a}(\omega)   
G_{SS12}^{r}(\omega)
  } \\
{\displaystyle  \times  \Bigl\lbrack n_F(\hbar \omega - e V +
\hbar \Omega/2) -  n_F (\hbar \omega + e V - \hbar \Omega/2) \Bigr\rbrack 
P(\Omega)  d \Omega  d \omega.} \end{array}  \label{30}
\end{equation}
Note that 
$P(\Omega)$ decreases when $\Omega$ becomes large.
In particular, the integrated contribution 
$\int_{\Omega}^{\infty} P(\Omega')  d\Omega'$
becomes negligible well {\it before} $\hbar \Omega>2eV$, 
allowing modify the upper bound of the integral over $\Omega$.
Simultaneously, the advanced (retarded) Green's functions are dressed 
as before and take the form :
\begin{equation}
G^{r,a}_{SS12}(\omega) = \frac{1}{D^{r,a}(\omega)}
\tilde{g}_{SS12}^{r,a}(\omega), \end{equation}
so that at zero temperature the current reads:
\begin{equation}
\langle I \rangle  \simeq  {16  e |\Gamma|^4 \over h \hbar W^4}  
\mid \! \Delta \! \mid^2   
\int_0^{2{eV \over \hbar} }  P(\Omega)  d \Omega \int_{0}^{{eV \over \hbar} - {\Omega \over 2}}  
{1 \over D(\omega)}  d \omega,
\label{31}
\end{equation}
with 
\begin{eqnarray}
D(\omega)  &=& \left({|\Delta|^2 \over \hbar^2} - \omega^2\right) D^r(\omega) D^a(\omega) = \left({|\Delta|^2 \over \hbar^2} - \omega^2\right) |D^a(\omega)|^2 \nonumber \\
&=& \left({\mid\! \Delta \! \mid^2 \over \hbar^2} - \omega^2 \right) 
\biggl( 1 + {|\Gamma|^4 \over W^4} \biggr)^2 
 +  4 {|\Gamma|^4 \over W^4} \omega^2. \label{32}
\end{eqnarray}
In a more rigorous approach, the expansions of $G^r_{SS12}(\omega)$ have
to be performed in the presence of phase fluctuations. This means that they
should contain not only linear terms in $P(\Omega)$ but also higher 
order correlators of the exponentiated phases. Here, in the expression for
 the current 
in Eq. (\ref{31}), only the first order in $P(\Omega)$ has been retained, which constitutes
the analog of a weak inelastic coupling -- single phonon -- 
approximation of Ref. \cite{combescot}. 

The  
differential conductance $d \langle I \rangle/dV$ can be straightforwardly
computed from Eq. (\ref{31}):
\begin{equation}
{d \langle I \rangle \over dV}  \simeq  {16 e^2 |\Gamma|^4 \over
h \hbar^2  W^4}  \mid \! \Delta \! \mid^2  \int_0^{2{eV \over \hbar}}  P(y)  {1
\over D\left({eV\over \hbar} - {y \over 2}\right)}  dy. \label{33} \end{equation}

\subsection{Finite frequency noise}
\label{frequency_noise}

Another useful tool which probes the fluctuations is 
the finite frequency noise. 
The symmetrized real time noise correlator  is given by:
\begin{equation}
S(t-t') = \langle \delta I(t)\; \delta I(t')
\rangle + \langle \delta I(t')\; \delta I(t) \rangle  
\label{36}
\end{equation}
where $ \delta I(t) = I(t) - \langle I(t) \rangle$.
This current--current correlator is expressed in terms of the
Keldysh Green's functions: using the self energy matrix, we can
write it as a trace.
\begin{eqnarray}
S(t,0)&=&-e^2 {\rm Tr}
\Big[\left\{\left(
\hat{\sigma}_z \hat{\Sigma}\hat{G}^{+-}_{SN}(0,t)
\hat{\sigma}_z\hat{\Sigma}\hat{G}^{-+}_{SN}(t,0)
-\hat{\sigma}_z\hat{\Sigma}^\dagger
\hat{G}^{+-}_{NN}(0,t)\hat{\sigma}_z\hat{\Sigma}
\hat{G}^{-+}_{SS}(t,0)\right)
+\left(~\hat{\Sigma}\leftrightarrow
\hat{\Sigma}^\dagger~;~S \leftrightarrow N~\right)\right\} \nonumber \\
&&~~~~~~~~~+ \left\{~ (0,t)\leftrightarrow (t,0)~ \right\} \Big],
\label{noise trace}
\end{eqnarray}

where $\hat{\sigma}_z$ is a Pauli matrix.

The lowest non vanishing contribution (order 4 in $\Gamma$) is
extracted by expanding the Green's functions in Eq. (\ref{noise
trace}). The corresponding Dyson equations read (time
integrations are implicit): 
\begin{eqnarray}
\hat{G}^{\pm\mp}_{SN}&=&\hat{G}^{r}_{SS}\hat{\Sigma}^\dagger
\hat{g}^{\pm\mp}_{NN}~,\nonumber\\
\hat{G}^{\pm\mp}_{NS}&=&\hat{g}^{\pm \mp}_{NN}\hat{\Sigma}
\hat{G}^{a}_{SS}~,\nonumber\\
\hat{G}^{\pm\mp}_{NN}&=&\hat{g}^{\pm\mp}_{NN}~,
\end{eqnarray}
where we have purposely written the Green's functions 
to the lowest order which they contribute. 
Inserting these expressions in the real time correlator gives:
\begin{eqnarray}
S(t,0)= 
-e^2\int_{-\infty}^{+\infty}  dt_1\int_{-\infty}^{+\infty} dt_2 {\rm Tr}
\Big[  \Big\{ ~&& 
\hat{\sigma}_z\hat{\Sigma}
\hat{G}^{r}_{SS}(0,t_1)
\hat{\Sigma}^\dagger
\hat{g}^{+-}_{NN}(t_1,t)
\hat{\sigma}_z\hat{\Sigma}
\hat{G}^{r}_{SS}(t,t_2)
\hat{\Sigma}^\dagger
\hat{g}^{-+}_{NN}(t_2,0)\nonumber\\
+&&
\hat{\sigma}_z \hat{\Sigma}^\dagger
\hat{g}^{+-}_{NN}(0,t_1)
\hat{\Sigma}
\hat{G}^{a}_{SS}(t_1,t)
\hat{\sigma}_z \hat{\Sigma}^\dagger
\hat{g}^{-+}_{NN}(t,t_2)
\hat{\Sigma}
\hat{G}^{a}_{SS}(t_2,0)\nonumber\\
-&&
\hat{\sigma}_z \hat{\Sigma}
\hat{G}^{r}_{SS}(0,t_1)
\hat{\Sigma}^\dagger
\hat{g}^{+-}_{NN}(t_1,t_2)
\hat{\Sigma}
\hat{G}^{a}_{SS}(t_2,t)
\hat{\sigma}_z\hat{\Sigma}^\dagger
\hat{g}^{-+}_{NN}(t,0)\nonumber\\
-&&
\hat{\sigma}_z\hat{\Sigma}^\dagger
\hat{g}^{+-}_{NN}(0,t)
\hat{\sigma}_z\hat{\Sigma}
\hat{G}^{r}_{SS}(t,t_1)
\hat{\Sigma}^\dagger
\hat{g}^{-+}_{NN}(t_1,t_2)
\hat{\Sigma}
\hat{G}^{a}_{SS}(t_2,0)
\Big\} + \Big\{ 0 \leftrightarrow t \Big\} \Big].
\label{noise with green}
\end{eqnarray}
The procedure for including the phase fluctuations in the noise
correlator is identical to that used in the expression of the
current of Eq. (\ref{current_K}). However, here additional
contributions which are proportional to 
$\hat{G}_{SS}^r\hat{G}_{SS}^r$ or to 
$\hat{G}_{SS}^a\hat{G}_{SS}^a$ occur, but these turn out not
contribute to the noise to lowest order in the phase
correlator.  

The current noise spectrum $S(\omega)$ is the Fourier transform
of $S(t)$. 
In the subgap regime, the only non--vanishing elements of the causal Green's
 functions are the off diagonal ones :
 $\tilde{g}^{a,r}_{SS12} =\tilde{g}^{a,r}_{SS21} \simeq
 \hbar/ W$. 
The current noise spectrum can be  expressed in terms of the  off 
diagonal Keldysh  Green's functions: 

\begin{eqnarray}
S(\omega) &\simeq& \frac{4 e^2}{\hbar^2} \frac{|\Gamma|^4}{2 \pi W^2} 
\int_0^{2{eV \over \hbar}}  d\Omega P(\Omega) \int_{-\infty}^{+\infty} d\omega_1 \nonumber \\
&&\Big[ g^{+-}_{N22} (\omega_1 - \Omega) g^{-+}_{N11} (\omega_1 -\omega)  \nonumber \\
&&+ g^{+-}_{N11} (\omega_1 + \Omega) g^{-+}_{N22} (\omega_1 -\omega) \nonumber \\
&&+ g^{+-}_{N11} (\omega_1 + \Omega) g^{-+}_{N22} (\omega_1 +\omega) \nonumber \\
&&+ g^{+-}_{N22} (\omega_1 - \Omega) g^{-+}_{N11} (\omega_1 +\omega) \Big].
\end{eqnarray}

The integration over $\omega_1$ defines the relevant 
energy intervals which contribute to the noise:

\begin{eqnarray}
S(\omega) &\simeq&  \frac{16 e^2}{2\pi} \frac{|\Gamma|^4}{W^4} 
\Big[ \int_{0}^{2{eV \over \hbar} -\omega} d\Omega 
P(\Omega)  \Theta\left({2eV \over \hbar} - \omega -\Omega\right) 
\left({2eV \over \hbar}  - \omega -\Omega\right)
+ \int_{0}^{2{eV \over \hbar} +\omega} d\Omega P(\Omega)   
\left({2eV \over \hbar} + \omega
-\Omega\right) \Big] \nonumber, \\
 \end{eqnarray}
 
where $\Theta(\Omega)$ is the Heaviside function. 

In the limit of low frequencies, the noise becomes:
\begin{eqnarray}
S(\omega=0) &=& \frac{32 e^2}{2 \pi} \frac{|\Gamma|^4}{W^4} 
\int_0^{2{eV \over \hbar}} d\Omega P(\Omega) \left({2eV \over
\hbar} - \Omega\right) \nonumber \\ &=& 4e \langle I \rangle, 
\end{eqnarray}

which corresponds to the Schottky formula. Note that this 
constitutes an illustration of Schottky formula for a situation where
the charge transfer is effectively inelastic: as shown in Fig.
\ref{rodolphe1}b, the transfer of two electrons in the
superconductor generates phase quanta and thus constitutes an
inelastic process.
 These processes are included when the
distribution $P(\omega)$ in the current and noise deviates from
a delta function.

 The current noise spectrum can be
measured experimentally and provides
  a direct way to obtain
information on $P(\omega)$ and thus on phase 
 fluctuations.
 In
the tunnel limit, the distribution $P(\omega)$ is related to
the second derivative of the current noise spectrum by the
formula  
 
\begin{equation}
 {d^2S \over d\omega^2}  \simeq 
\frac{\hbar \langle I \rangle_0}{V} 
 \left[ P\left({2eV \over \hbar} + \omega\right) +
P\left({2eV \over \hbar}-\omega\right)\Theta\left({2eV \over \hbar}-\omega\right) \right]. 
 \label{37}
\end{equation}
 
The prefactor stands for the current in the absence of phase fluctuations :$
\langle I \rangle_0 = (16 e^2 V |\Gamma|^4) /  (\hbar W^4)$.

Note that so far, no specific model for the phase fluctuations 
has been specified, as the only assumption used what the fact that 
$P(\Omega)$ falls sufficiently rapidly under the gap. 
In the next section, current voltage characteristics are plotted 
using a specific model to describe the phase dynamics.

\section{Application to the 1D Josephson-Junction Array model}

\label{section5}

One spatial dimension is specified, motivated by the fact that
tunneling typically occurs near
 the tip of a scanning tunneling
microscope. The dynamics of the phase can then be modeled by a
one-dimensional Josephson-junction Array (JJA), which was 
previously studied in linear response \cite{falci}.  
The phase correlator in Eq. (\ref{phase_correlator_a})
is then known. The parameters of this model are the 
Josephson energy $E_J$ and the charging energy $E_0$ (which in
turn define the plasma frequency), together  with the parameter
$\kappa$ which depends  on the ratio of the
junction capacitance divided by  the capacitance of the islands.
If the latter capacitance dominates over that of the junction, 
$\kappa=2$. 

The importance of phase fluctuations is monitored by the
ratio $\hbar\omega_p/E_J$.
In particular, for $\hbar\omega_p/E_J< \pi/2$, 
the phase Hamiltonian can be mapped to a  Luttinger liquid
model and $P(\Omega)$ 
can be derived in a standard way
from harmonic oscillator correlators.
On the other hand, for $\hbar\omega_p/E_J>\pi/2$, 
the phase correlator decays 
exponentially in time.
This is identified as the disordered phase.
At $\hbar\omega_p/E_J=\pi/2$, there is a  
Kosterlitz-Thouless transition between the ordered and the 
disordered phase.

\subsection{Ordered phase}

At zero temperature, the time dependent phase correlator is given
by:
 
\begin{equation}
 p(t)  =  \exp
\Biggl\lbrace - {\sqrt{2} \over \pi} 
   \sqrt{ {E_0 \over E_J}}
 \Bigl\lbrack
  \int_0^{\kappa \omega_p \mid t\mid } { 1 - \cos  x
\over x}  {dx \over \sqrt{1-(x/\kappa \omega_p t)^2}} +  
 i\; {\rm sgn}(t) \int_0^{\kappa \omega_p \mid t\mid } {\sin  x
\over x}  {dx \over \sqrt{1-(x/\kappa \omega_p t)^2}} 
  \Bigr\rbrack  \Biggr\rbrace
~,  \label{34}
\end{equation}
with ${\rm sgn}(t)$ the sign function.
Its Fourier transform $P(\Omega)$ is identically zero for
$\Omega<0$.
 For $\Omega>0$, it has a power law behavior 
close to $\Omega=0$:
 
\begin{equation}
P(\Omega)  \simeq
 \Omega^{-\bigl( 1 - {1\over\pi} \sqrt{{2 E_0 \over E_J}} \bigr)
}, \label{35} 
\end{equation}
and goes rapidly to zero for $\Omega>\kappa\omega_p$.

For $\Omega = \kappa \omega_p$, $P(\Omega)$ shows a weaker 
singularity \cite{falci} and diverges as $( \omega_p - \Omega
)^{-\alpha}$, with  $\alpha=1/2-\pi^{-1}\sqrt{2E_0/E_J}$ (here
this corresponds to fixing the parameter of Ref. \cite{falci}
$\kappa=2$). 

Turning to the transport properties, the effect of the phase
fluctuations on the noise are probed. For convenience, we compare
our results to those derived with a BTK model\cite{BTK}, which
corresponds to an interface with a  rigid superconductor. 
The quantity $\gamma=\Gamma/W$ corresponds to the transparency
of the barrier. $\gamma$ enters the prefactor of the delta
function potential of BTK theory
with the dependence $\hbar^2k_F(1-\gamma^2)/2m\gamma$ ($k_F$ is
the Fermi wave vector). $\gamma=1$ corresponds to a perfect
contact, while $\gamma\to 0$ describes an opaque barrier. 

\begin{figure}  
\epsfxsize 8 cm  
\centerline{\epsffile{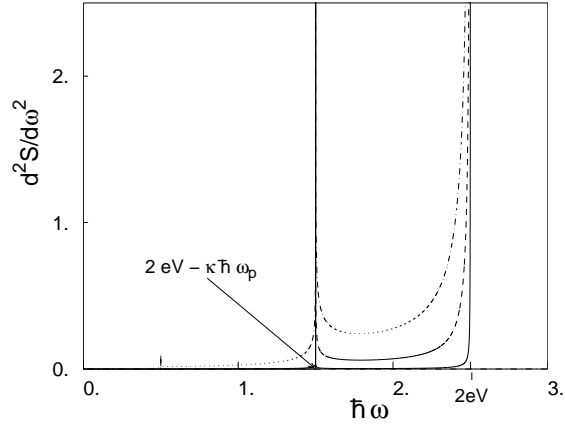}}  
\medskip  
\caption{\label{pierre2}  Second derivative of the symmetrized
noise spectrum versus frequency (in units of
$4(e^2/h)\hbar\gamma^4/\omega_p$) for different
  values of $\hbar\omega_p / E_J$, $\hbar\omega_p / E_J=0.01$
(solid line), $\hbar\omega_p / E_J=0.2$ (dashed line), $\hbar\omega_p / E_J=1$ (dashed dotted
line).
 The plasma frequency $\omega_p$ is set to $|\Delta|/2\hbar$.}  
\end{figure} 

In Fig. \ref{pierre2} the second derivative of the symmetrized
noise spectrum  $d^2S / d \omega^2$ is plotted as a function of
$\omega$, for a perfectly good contact. Several values of the
parameter $\hbar\omega_p/E_J$ are
 considered, starting from weak
fluctuations to substantial
 collective excitations within the
ordered phase. To illustrate the effect of phase fluctuations,
the choice $\hbar \omega_p=|\Delta|/2$ is made, so that all the
features in   $P(\Omega)$ are located  
within the superconducting gap.

In the absence of phase fluctuations, the finite frequency noise
has a singular derivative at $\hbar\omega=2eV$ \cite{torres}. In Fig.
\ref{pierre2}, this would imply a delta function peak in $d^2S
/ d \omega^2$. 
For low voltages, there is no deviation with respect
to the results of scattering theory.
The effect of phase fluctuations is twofold: first,
the delta function broadens to a power law divergence 
(partially cut on the figure), and 
second, it acquires a secondary peak -- previously
discussed for $P(\Omega)$ -- at $2eV-\kappa\hbar \omega_p$. 
This illustrates
the result of Eq. (\ref{37}): an experimental measurement of the 
current noise spectrum constitutes a direct diagnosis of the 
importance of phase fluctuations.

Naturally, the differential conductance $d\langle I \rangle / dV$
in Eq. (\ref{33})
is also affected by the phase fluctuations. It is studied below 
for different regimes: first, it is plotted as a function of
bias, for both the case of a perfect contact and for the case of
a weak transmission; second, the transparency is varied while the
bias is fixed to a large value in order to explore the
deviations from the tunnel limit. In each case, curves are
obtained for several values of the ratio $\hbar \omega_p/E_J$ which 
characterizes the importance of the fluctuations.  

\begin{figure}  
\epsfxsize 8 cm  
\centerline{\epsffile{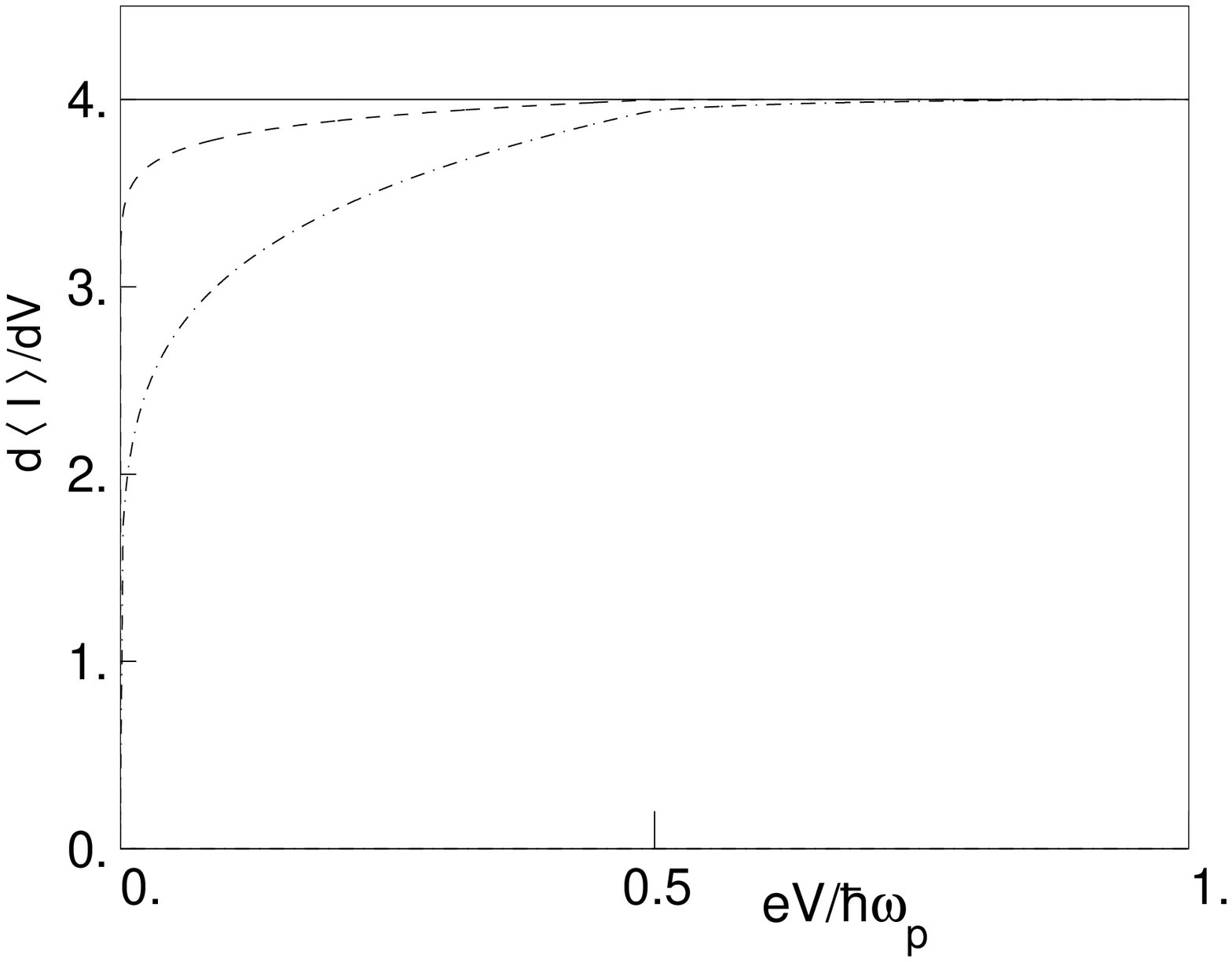}}  
\medskip  
\caption{\label{pierre3}  Differential conductance $d \langle I \rangle / dV$
 in units of $e^2 / h$ versus the ratio $eV /\hbar
\omega_p$. A perfect contact 
 ($\gamma = 1$) is considered and
different values of $\hbar\omega_p / E_J$ are displayed : $\hbar\omega_p / E_J=0$ (BTK curve,
solid  line), 
 $\hbar\omega_p / E_J=0.2$ (dashed line) and $\hbar\omega_p / E_J=1$ (dash-dotted
line).}   
 \end{figure}

The differential conductance of a perfect junction is displayed
in Fig. \ref{pierre3} as a horizontal line at 
$d\langle I \rangle / dV=4e^2/h$ \cite{BTK}.
In the presence of fluctuations, it saturates to the 
BTK value for biases larger than $\hbar \omega_p/2$
(here the bias is varied from $0$ to $|\Delta|$):
for sufficiently large $eV$,  $\int_0^{2eV/\hbar}P(\Omega)d\Omega$ is
essentially equal to unity because for $\gamma = 1$, 
$D(x)$ which enters Eq. (\ref{33}) is a constant.
Phase fluctuations have a tendency to decrease the differential
conductance at low bias. In particular for $\hbar\omega_p/E_J\geq 1$,
the significant contributions from the integral 
$\int_{\omega_p}^{2eV/\hbar}P(\Omega)d\Omega$ are the cause for the 
deviations from the ideal conductance at large bias. 

\begin{figure}  
\epsfxsize 8 cm  
\centerline{\epsffile{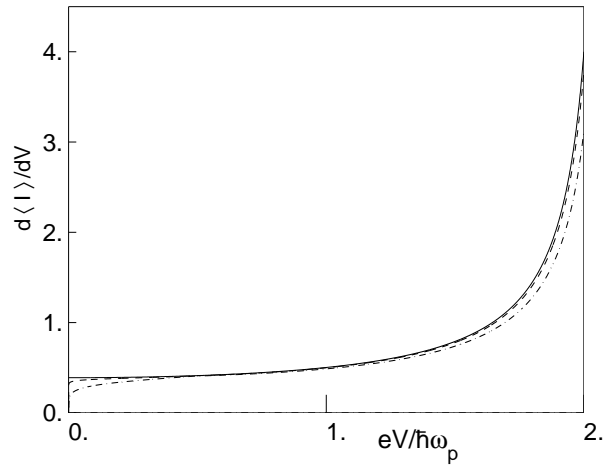}}  
\medskip  
\caption{\label{pierre4}  Same as Fig. \ref{pierre3}, for
a transparency  $\gamma=0.4$.  }   
\end{figure}

The case of $\gamma=1$ is somewhat
academic because perfect contacts are difficult to achieve in
practice. 
The same curves are plotted now for $\gamma=0.4$
in Fig. \ref{pierre4}. For intermediate biases, 
phase fluctuations do not affect the
differential conductance drastically: at $eV= \hbar\omega_p$ for
instance, $d \langle I \rangle / dV$ is essentially
constant (of the order $4e^2\gamma^2/h$), and all curves can not
be distinguished. However, deviations occur both at small and
large voltages. For small voltages, according to Eq.
(\ref{33}) and due to the fact that $P(0)=0$ the
differential conductance in the presence of {\it any}
phase fluctuation is required to vanish (this is not really
obvious in Fig. \ref{pierre4} because of the choice of scale).
More dramatic is the fact that phase fluctuations 
play a role at large voltages, contrary to the high
transparency regime.   

\begin{figure}   \epsfxsize 8 cm  
\centerline{\epsffile{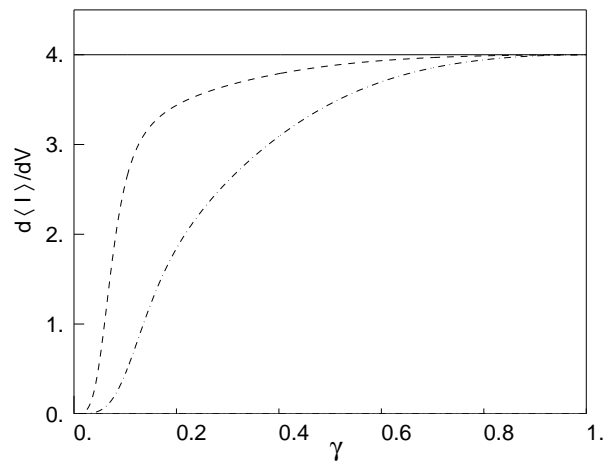}}   \medskip 
 \caption{\label{pierre5}
 Differential conductance
(in units of $e^2 / h$) as a function of the transparency,
for a voltage  bias $eV = |\Delta|$ (same convention for
$\hbar\omega_p / E_J$ as in Fig. 3)} 
\end{figure}

In Fig. \ref{pierre5}, the transparent regime is examined 
with a bias voltage $eV = |\Delta|$. In contrast to the
previous curves, the differential conductance is plotted versus
the transparency of the barrier. The
plasma frequency and the ratio $\hbar\omega_p/E_J$ are chosen as
previously.   

For barrier transparencies smaller than $0.1$, the
differential conductance has power law dependence (${d \langle I
\rangle  / dV}\sim \gamma^4$). As the transparency is
increased, in the presence of phase
fluctuations an inflexion point appears and the rise of ${d
\langle I \rangle  / dV}$ is then slower. Note once again that
it is reduced from the BTK case when phase fluctuations
become important. As indicated previously in Fig. \ref{pierre3},
the differential
conductance saturates to the BTK value at high $\gamma$ regardless
of the degree of phase fluctuations.    

\subsection{Disordered phase}  

In the disordered phase ($\hbar\omega_p/E_J >\pi/2$), the behavior
of the Fourier transform of the time dependent phase correlator
close to the transition  has been previously derived from the
$XY$ model \cite{heinekamp}. For small $\Omega$,     
\begin{equation}
P(\Omega) \sim  \Theta(\Omega - \omega_p \xi^{-1})    (\Omega -
\omega_p \xi^{-1})^{-1/2}, \label{kosterlitz}
\end{equation}  
where $\Theta(x)$ is the Heaviside function.   The results of
linear response theory of Ref. \cite{falci} are briefly recalled
and compared to the predictions of the Keldysh calculation
of Eq. (\ref{31}). In this phase,
a threshold voltage $\hbar \omega_p  \xi^{-1}$ has to be reached in
order to obtain a non zero current. Here, the
Kosterlitz-Thouless correlation length (in dimensionless
units) reads $\xi = \exp \left[b \left(
2/ \pi - \sqrt{E_J/8 E_0} \right)^{-1/2}\right]$, where
 $b$ is a positive constant.
Because the goal is  to predict the current as
a function of voltage  for both the tunnel limit and the
transparent regime (for $0 < eV <  \mid\!\Delta\!\mid$), 
$P(\Omega)$ needs to be characterized.  
$P(\Omega)$ can be set to zero for $\Omega$ larger than
$\kappa\omega_p$. This assumption is reasonable below the
Kosterlitz-Thouless transition.  In accordance with 
Eq. (\ref{kosterlitz}), we adopt the form  
\begin{equation}
P(\Omega)  \simeq  {1\over 2}((\kappa -
\xi^{-1})\omega_p)^{-1/2}   \Theta(\Omega - \omega_p
\xi^{-1})   \Theta (\kappa\omega_p - \Omega)  (\Omega- \omega_p
\xi^{-1})^{-1/2}, \label{kosterlitz2}  \end{equation} 
Using Eq. (\ref{31}) and (\ref{kosterlitz2}), calculations
can be performed analytically.

For $eV < \hbar \omega_p \xi^{-1}/2$, the average current
vanishes.  For $\xi^{-1}  < 2eV/ \hbar \omega_p  < \kappa$, the
average current reads :  
\begin{equation} \langle I \rangle  =  
{e \over h \hbar}  {16 \gamma^4 \over (1 + \gamma^4)^2}   
{\kappa\omega_p\over2}   (1-(\kappa\xi)^{-1}) \left( \frac{2
{eV\over \hbar \omega_p} - \xi^{-1}}{ \kappa-\xi^{-1}}
\right)^{3/2}  f({\cal A}), \label{44} \end{equation} where
\begin{mathletters}
\begin{eqnarray}
f(x)  &=&  {2\over x} \Big\{ {1\over2} \sqrt{1+{2\over x}} \; \ln \Big| \frac{1+
  \sqrt{1+{2\over x}}}{1-\sqrt{1+{2\over x}}}\Big| - \sqrt{{2\over x}-1} 
\; \arctan \left(\frac{1}{\sqrt{{2\over x}-1}}\right) \Big\} ~,
\\ {\cal A}  &=&  \left({2eV \over \hbar\omega_p}  -
\xi^{-1}\right)  \biggl( {\hbar \omega_p \over
\mid\!\Delta\!\mid} \biggr)   \sqrt{1 - {4 \gamma^4 \over (1 +
\gamma^4)^2 }}~, \label{scal_var} \end{eqnarray}
\end{mathletters}
where ${\cal A}$ is a scaling variable which varies between $0$
(perfect contact $\gamma=1$) and $1$ ($\gamma=0$  and a plasma
frequency $\omega_p = |\Delta|/\hbar$). Note that the function 
$f$ does not vary substantially between these two 
transparencies: $f({\cal A}=0)=2/3$ and $f({\cal A}=1)\simeq
0.71$. The dependence of the current on the parameters $\gamma$, 
$eV$, $\omega_p$ and $\xi$ is given by the prefactor in Eq.
(\ref{44}).   

\begin{figure}  
\epsfxsize 8 cm  
\centerline{\epsffile{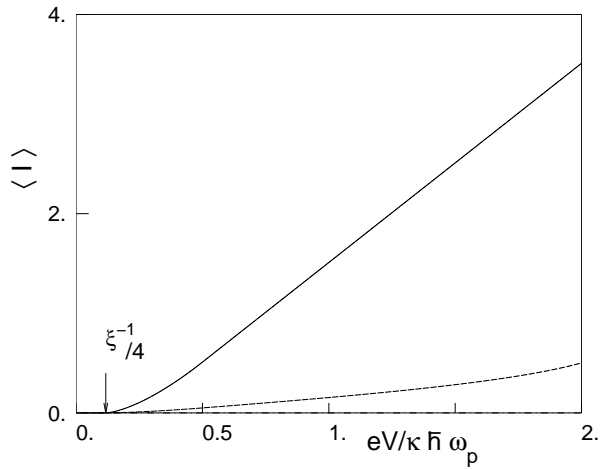}}  
\medskip  
\caption{\label{pierre6}    Current $\langle I \rangle$ versus
bias voltage in the disordered phase for $\hbar\omega_p / E_J=2
\pi$ [in units of $(4e\omega_p/\pi)\gamma^4/(1+\gamma^4)^2$]:
perfectly good contact (solid line)  and for $\gamma=0.4$
(dashed line). }    \end{figure}

For $\kappa\hbar \omega_p/2 < eV < \mid\!\Delta\!\mid$, 
\begin{equation}
\langle I \rangle  =   {e \over h \hbar}  
{16 \gamma^4 \over (1 + \gamma^4)^2}  
 {\omega_p\over 2}    (\kappa - \xi^{-1}) g\left({\cal A},
\left( {2eV \over \hbar \omega_p}
 - \xi^{-1}\right)  (\kappa - \xi^{-1})^{-1}
\right), \label{46}
\end{equation}
where
\begin{eqnarray}
g(x,y)  &=&  \Bigl({y\over x}\Bigr)
\Biggl( \ln \left\vert 1 +{x\over 2} \left( 1 -{1 \over y}
\right) \right\vert -  \ln \left\vert 1 -{x\over 2}
\left( 1-{1\over y} \right) \right\vert \nonumber \\
  && ~~~~  -
\sqrt{{y(2+x) \over x}}  \biggl\lbrace \ln \biggl\vert
1-\sqrt{{x \over (2+x)y}} \biggr\vert   - \ln \biggl\vert
1+\sqrt{{x \over (2+x)y}} \biggr\vert  \biggr\rbrace  -  2
\sqrt{{y(2-x) \over x}}   \arctan \Biggl(\sqrt{{x \over
y(2-x)}} \Biggr) \Biggr) , \label{47} 
\end{eqnarray}
For $\gamma=1$, the behavior of $\langle I \rangle$ is
essentially the same as that of linear theory \cite{falci}, 
for arbitrary  $\omega_p$. These results are plotted
in Fig. \ref{pierre6} (with the choice  
$\omega_p=\Delta/4\hbar$).

For $eV$ larger than $\kappa\hbar \omega_p/2$, the curve 
$\langle I \rangle$  versus $eV$ is a straight line.
For less transparent barriers ($\gamma \not= 1$), 
two behaviors have to be distinguished.
First, if the plasma frequency
is smaller than the gap parameter, then the scaling
variable ${\cal A}$ in Eq. (\ref{scal_var}) is small and linear
response is again retrieved. Second, if $\hbar \omega_p$ is
comparable to $\mid\!\Delta\!\mid$,  $\langle I \rangle$
increases typically as $(2eV - \kappa\hbar \omega_p
\xi^{-1}/2)^{3/2}$ for $\xi^{-1}/2< 2eV/\kappa\hbar \omega_p <
1$. Then, for $eV$ larger than $\kappa\hbar \omega_p/2$, the
response is non-linear.  In particular, when the transparency is
low, $\langle I \rangle$ increases faster
than  $(2eV/\kappa\hbar\omega_p - \xi^{-1}/2)^{3/2}$ for $eV >
\kappa\hbar \omega_p/2$. Overall, no sensible deviations from
linear response are  expected in the disordered limit.

\section{Conclusion}

A formulation of quantum transport in normal
metal superconductor junctions is developed in the case where
the order parameter undergoes phase fluctuations, both in the
tunnel limit and in the transparent regime. 
The Keldysh Green's function was used, allowing to go
beyond linear perturbation theory, provided that the subgap
regime is specified (this latter assumption enabled us to
neglect some contributions in the Dyson expansion).  While we
applied  our results to the 1D Josephson junction array model of
Ref. \cite{falci}, the result is quite general as any model 
for the phase correlator can be employed to calculate the
current and the finite frequency noise. 
An interesting issue is that the zero frequency noise satisfies
the Schottky formula (as should be), with a current describing
inelastic processes. A direct correspondence between the phase
correlator  and the second derivative of the noise spectral
density with respect to frequency has been established. For a
highly transparent interface, phase fluctuations  are shown to
affect only the differential conductance at low voltages: at
high voltage the $I(V)$ characteristic is essentially that of a
rigid superconductor, which can be described with the BTK
model. On the other hand, for a less transparent barrier,
deviations with respect to the BTK are shown to occur both at
high and low voltages. This effect becomes more dramatic when
the barrier is opaque.  

The present system bears similarities with other situations
where Andreev reflection is accompanied by the emission of 
a bosonic excitation. As mentioned in the introduction, 
Ref. \cite{falko} dealt with a ferromagnet whose magnons 
can flip the spin of one of two same spin electrons
in order to contribute to the subgap conductance
in a s-wave superconductor.
Therefore, electrons with opposite spin contribute 
to the current as usual, but a new scattering 
channel -- due to this one-magnon process -- is opened, and
enhances the Andreev current overall. 
The contrast with our calculation is
that here, the current is always reduced by the excitation
of phase modes.  For a specified voltage bias, the more phase
fluctuations, the more energetic are the modes which propagate
in the Josephson junction array. On the side of
the normal metal, this requires that the sum of the
energies of the electron and hole has to be
large. Yet the Fermi level restricts the 
impinging electron energy. This reduction of the available
phase space for inelastic assisted Andreev reflection
thus justifies a weaker current when the fluctuations 
are large. Also, note that the phase correlator which was used
in this calculation typically appears in situations where 
a quantum mechanical degree of freedom is coupled to an
electromagnetic environement \cite{devoret_grabert}. Overall,
if one includes multi-magnon processes to the situation of
Ref. \cite{falko}, both  ferromagnetic/superconductor  and
normal-metal/fluctuating superconductor junctions 
constitute illustrations of quantum dissipative systems
\cite{weiss}, where the quantum process to be probed
-- Andreev reflection --  is coupled
to a physical environment (a bath of magnons/``phase
oscillators'').             

Potential applications to physical 
systems are now addressed.
While typical BCS superconductors have  essentially amplitude
fluctuations  (except close to the
superconducting transition), the  high-$T_c$ cuprates  have a
much lower phase stiffness and hence, are expected to show 
strong fluctuations of the phase of the order parameter
\cite{emery}. 
Here, an s-wave order parameter was assumed for
simplicity, whereas these materials are mainly d-wave, but our
calculation could in principle be generalized to this latter
situation. 
The two main
differences between s-wave and d-wave materials are 
the following: first, in a d-wave material,  there are nodes in
the order parameter in specific directions, which results in a
finite density of states for
quasiparticles near the superconductor chemical potential.
Second, low energy Andreev bound states may exist at this energy.
Depending on the orientation of momentum, an
electron-like quasiparticle and a hole-like
quasiparticle in the superconductor  see 
an  order parameter with the opposite sign \cite{tanaka}. 
This generates a zero-bias conductance peak (ZBCP) which is a
hallmark of a   d-wave  order parameter. 

At first sight, and for ideal interfaces, the connection of
between our s-wave model and a d-wave compound could be made by
first choosing  the antinodal direction   to be orthogonal to the
plane of the interface \cite{sheehy}, second by restricting
electron transfer along the normal of this interface.
Nevertheless, the confrontation of our theoretical results with
transport experiments in high-$ T_c$ compounds may prove
too difficult, as surface defects are likely to be present in
such compounds. Such defects are responsible for a 
reduction of the  order parameter 
in the vicinity of the junction. Moreover, 
the presence of 
Andreev bound states -- also induced by the impurities at the
interface -- renders the  present approach inadequate. 
Furthermore, surface roughness provokes a randomization of the
tunneling directions so that the above scenario for antinodal
tunneling loses its practicality.   

Previous theoretical works addressing the role
of phase fluctuations \cite{dorsey} on Andreev reflexion in the
pseudogap phase \cite{sheehy,choi}  focussed on the mildly
underdoped case: the hole density is chosen on the
right hand side of the zero temperature insulator-superconductor
quantum critical point (QCP). In this regime quantum
fluctuations may renormalized  the superfluid density but
the physics turns out to be essentially that of the
two-dimensional classical XY model (renormalized classical regime
in the spirit  of the $n=2$, non-linear $\sigma$-model). 
Transverse  superfluid velocity fluctuations due to the motion
of vortices are expected to dominate. Because the time scales
for the vortex motion, as measured by THz spectroscopy
\cite{corson}  are much larger than the Andreev time $\tau_A =
\hbar/\Delta$, the correlation of the phase  was treated as
static (on the scale of $\tau_A$). The phase correlator decays
in space over a   lengthscale $\xi_{cl}$ which is the
classical correlation length \cite{bkt}.   If tunneling
occurs on an area which is much smaller than  $\xi_{cl}$, 
then an Andreev signal is expected. 

Our
calculation (albeit in 1 dimension), focuses on quantum
fluctuations. Fluctuations are included in a non-perturbative and
systematic manner, regardless of the phase propagator   which
is used in the end. As a general trend, quantum fluctuations
were found to have a tendency to reduce the Andreev reflexion
signal.  Note that other models of pseudogap also rely
partly on quantum fluctuations \cite{balents}, while others  
incorporate aspects of 1D physics (stripes) \cite{salkola}. 
Extensions of our calculation  beyond 1D
-- together with surface roughness --
or with  competing orders could be envisioned to 
achieve a more realistic description of  
STM experiments on high-$T_c$ compounds. 

\acknowledgements  
We thank G. Falci for clarifications about the phase
correlator in the JJA model. Discussions with C. Bruder, D.
Feinberg and F. Hekking are gratefully acknowledged.


\begin{thebibliography}{10} 

\bibitem{bruder} C. Bruder,
Phys. Rev. B {\bf 41}, 4017 (1990).

\bibitem{tanaka} Y. Tanaka and
S. Kashiwaya, Phys. Rev. Lett. {\bf 74}, 3451 (1995);   S.
Kashiwaya, Y. Tanaka, M. Koyanagi, and   K. Kajimura, Phys.
Rev. B {\bf 51}, 1350 (1995).

\bibitem{kim} Y. B. Kim and X. G. Wen, Phys. Rev. B {\bf
48}, 6319 (1993).

\bibitem{falci} G. Falci, R. Fazio, A. Tagliacozzo, and
 G. Giaquinta, Europhys. Lett. {\bf 30}, 169 (1995).

\bibitem{sheehy} D. E. Sheehy, P. M. Goldbart, J.  Schmalian, and
A. Yazdani,  Phys. Rev. B {\bf 62}, 4105 (2000).

\bibitem{choi} H. Y. Choi, Y. Bang, and D. K. Campbell, Phys.
Rev. B {\bf 61}, 9748 (2000);  
Y. Bang and H. Y. Choi, Phys. Rev. B {\bf 62}, 11763 (2000).

\bibitem{caroli} C. Caroli, R. Combescot, P. Nozi\`eres, and
D. Saint-James,  J. Phys. C {\bf 4}, 916 (1971).

\bibitem{martin-rodero} A. Mart{\'\i}n-Rodero,
A. Levy Yeyati, and F. J. Garc{\'\i}a-Vidal, Phys. Rev. B
{\bf 53}, R8891 (1996). J. C. Cuevas, A. Mart{\'\i}n-Rodero, and
A. Levy Yeyati,  Phys. Rev. B {\bf 54}, 7366 (1996).

\bibitem{falko} E. McCann and V.I. Fal'ko, Europhys. Lett. {\bf
56}, 583 (2001); G. Tkachov, E. McCann and V.I. Fal'ko, Phys.
Rev. B {\bf 65}, 024519 (2001).

\bibitem{deutscher} G. Deutscher, Nature {\bf
397}, 410 (1999);  Y. Dagan, A. Kohen, G. Deutscher, and A.
Revcolevschi, Phys. Rev. B {\bf 61}, 7012 (2000).

\bibitem{gonnelli} R. S.  Gonnelli {\it et al}, 
cond-mat 0101209 (2001).

\bibitem{mourachkine} A. Mourachkine, Europhys. Lett. {\bf
50}, 663 (2000).

\bibitem{De Palo} S. De Palo, C. Castellani, C. Di Castro, and
B. K. Chakraverty,  Phys. Rev. B {\bf 60}, 564 (1999).

\bibitem{glazman}
L. G. Glazman. and A. I. Larkin,
 Phys. Rev. Lett. {\bf 79}, 3736 (1997).

\bibitem{ketterson}
J. B. Ketterson and S. N. Song,
{\it Superconductivity} (1999) Cambridge University Press.

\bibitem{bkt} V. L. Berenzinskii, Zh. Eksp. Theor. Fiz. {\bf
59}, 907 (1970) [Sov. Phys. JETP {\bf 32}, 493 (1971)]; J. M.
Kosterlitz, and D. J. Thouless, J. Phys. C {\bf 6}, 1181 (1973). 
%
\bibitem{heinekamp} S. W. Heinekamp and R. A. Pelcovits, 
Phys. Rev. B {\bf 32}, 4528 (1985).
%
\bibitem{combescot} C. Caroli, R. Combescot, P. Nozi\`eres and D. Saint-James,
J. Phys C {\bf 5}, 21 (1972).

\bibitem{BTK} G. E. Blonder, M. Tinkham, and T. M. Klapwijk,
Phys. Rev. B {\bf 25}, 4515 (1982).

\bibitem{torres} J. Torr\`es, T. Martin and G.B. Lesovik, Phys.
Rev. B {\bf 63}, 134517 (2001).
%
 
\bibitem{devoret_grabert} 
G. L. Ingold and Yu. V. Nazarov, in {\it Single charge
tunneling}, H. Grabert and M. Devoret Editors 
(plenum Press, New York 1992).

\bibitem{weiss} U. Weiss, {\it Quantum dissipative systems},
(World Scientific, Singapore 1993).

\bibitem{emery} V. J. Emery and S. A. Kivelson, Nature (London) {\bf 374}, 434 (1995).

\bibitem{dorsey} H. Y. Kwon and A. T. Dorsey, Phys. Rev. B {\bf 59},
 6438 (1999).

\bibitem{corson} J. Corson, R. Mallozzi, J. Orenstein, J. N. Eckstein, 
and I. Bozovic, Nature (London) {\bf398}, 221 (1999).

\bibitem{balents} L. Balents, M. P. A. Fisher, and C. Nayak, Int. J. Mod. Phys. B {\bf 12}, 1033 (1998).

\bibitem{salkola} M. Salkola, V. J. Emery, and S. A. Kivelson, Phys. Rev. Lett. {\bf 77}, 155 (1996).


\end{thebibliography}
\end{document}